\documentclass[aps,floats,showpacs]{revtex4}
\usepackage{amsmath}

\usepackage{epsfig}
\usepackage{graphicx}
\usepackage{dcolumn}
\usepackage{bm}

\def\gapp{\lower.35em\hbox{$\stackrel{\textstyle>}{\sim}$}}
\def\lapp{\lower.35em\hbox{$\stackrel{\textstyle<}{\sim}$}}

\begin{document}
\bibliographystyle{apsrev}
%


\title{Magnetic moments in the presence of topological defects  in graphene}
\author{Mar\'{\i}a P. L\'opez-Sancho, Fernando de Juan, and Mar\'{\i}a A. H. Vozmediano}
\affiliation{Instituto de Ciencia de Materiales de Madrid,\\
CSIC, Cantoblanco, E-28049 Madrid, Spain.}

\date{\today}
\begin{abstract}
We study the influence of pentagons, dislocations and other
topological defects breaking the sublattice symmetry on the
magnetic properties of a graphene lattice in a Hartree Fock mean
field scheme. The ground state of the system with a number of
vacancies or similar defects belonging to the same sublattice is
known to have  total spin equal to the number of uncoordinated
atoms in the lattice for any value of the Coulomb repulsion U
according to the Lieb theorem. We show that the presence of a
single pentagonal ring in a large lattice is enough to alter this
behavior and a critical value of U is needed to get the polarized
ground state. Glide dislocations made of a pentagon-heptagon pair
induce more dramatic changes on the lattice and the critical value
of U needed to polarize the ground state depends on the density
and on the relative position of the defects. We found a region in
parameter space where the polarized and unpolarized ground states
coexist.

\end{abstract}
%
%
%
%
 \maketitle


The recent synthesis of a single layer of graphite
\cite{Netal05,ZTSK05} and the expectations of  future
nanoelectronics applications has renewed the interest in graphitic
materials. Among the possible exotic properties of graphene,
magnetism is one of the least studied and most appealing given the
interest of potential applications of organic magnets.
Ferromagnetic order enhanced by proton irradiation has been
observed in graphite samples \cite{Betal07} and demonstrated to be
due to the carbon atoms by dichroism experiments \cite{Oetal07}.
Ferromagnetism has also been reported in carbon nanotubes induced
by magnetic impurities \cite{MF06} and in honeycomb lattice
arranges of first row elements \cite{OO01}. By now it is clear
that the underlying mechanism leading to ferromagnetism in all
carbon structures is the existence of unpaired spins at defects
induced by a change in the coordination of the carbon atoms
(vacancies, edges or related defects)\cite{KM03} although the
mechanism for the occurrence of long range magnetic order is still
unknown. The magnetic properties of vacancies and voids  in
graphene layers and in graphene nanoribbons have been investigated
by different techniques
\cite{Fetal96,Wetal99,MNFL04,PGetal06,PMH08,CPetal08} and in all
cases the spin of the defects and the magnetic ordering was
determined by the sublattice imbalance.

Graphene is made of carbon atoms arranged in a two dimensional
hexagonal lattice that can be seen as two interpenetrating
triangular lattices A and B. It is the peculiar geometric
structure of the honeycomb lattice with two atoms per unit cell
what determines the very interesting low energy properties of the
system whose quasiparticles are massless Dirac fermions in two
dimensions \cite{Netal05}. The graphene lattice is an example of a
{\it bipartite} lattice: it is made of two sets of sites A and B
and the coordination is such that atoms of either set are only
connected to atoms belonging to the opposite subset. In a
beautiful paper concerning the magnetic properties of the Hubbard
model in bipartite lattices, E. Lieb \cite{L89} proved  a theorem
stating that for a repulsive value of the Hubbard interaction U
the ground state of the half filled lattice is non degenerate and
has a total spin equal to half the number of unbalanced atoms:
$2S=N_A-N_B$. This rule has been confirmed recently in a number of
studies of graphene with vacancies, edges or larger defects
\cite{VLSG05,KH07,PLC08,PFB08} and the Lieb theorem has become a
paradigm of magnetic studies in graphene clusters and in
nanographite. What is more interesting, the rule seems to survive
when more complicated calculations such as {\it ab initio},
density functional or molecular dynamics are performed. The
purpose of this letter is to emphasize the fact that the crucial
property that determines the magnetic behavior of the lattice is
its bipartite nature as it was already established in the original
paper \cite{L89}. Vacancies, islands, cracks or whatever defects
preserving this property will in most cases obey the Lieb rule. We
will show that a slight frustration of the bipartite property
(known as sublattice symmetry in the graphene community) is enough
to alter the rule.

The importance of the findings presented in this work  is not
purely academic: pentagonal or heptagonal rings, dislocations,
Stone Wales, and other topological defects breaking the sublattice
symmetry will most probably be present in the graphene samples and
hence the magnetic properties predicted for graphene system,
nanoribbons or small clusters should be revised.

Recent observations of extraordinary mechanical stiffness
coexisting with ripples in large graphene samples \cite{BBetal08}
points towards topological defects as the main source of curvature
\cite{CV07a,CV07b} as it is known that elasticity can not give
rise to curvature in two dimensions \cite{SN88}. Nucleation of
dislocations in the fabrication of the samples by mechanical
cleavage of graphite is practically unavoidable and they should be
observed in local  probes as scanning tunnelling microscopy.
Direct observation can be hard due to their very local influence
on the electronic properties \cite{Cetal08} and the present work
provides alternative indirect ways to detect the presence of
dislocations of other topological defects through their influence
on the magnetic properties. Although the theoretical description
of topological defects in the continuum limit was set some time
ago \cite{GGV92,GGV93,GGV01} and their influence  on the
electronic and transport properties of graphene has been studied
recently in a number of papers \cite{CV07a,CV07b,CV07c,GHD08,G08},
their implications on the magnetic structure has not been fully
explored.

 We show that the ground state of the honeycomb lattice in the
presence of pentagonal, heptagonal rings or dislocations deviates
from the predictions of the Lieb theorem. In the classical
configuration of a graphene lattice with several vacancies of the
same sublattice the total spin of the ground state is half the
number of unpaired sites for any arbitrarily small value of the
Hubbard repulsion U. This behavior is due to the presence of zero
energy states generated by the unpaired electrons in the bipartite
lattice and its subsequent polarization when an electron-electron
interaction U is added.

In the presence of the topological defects discussed in this work
a finite critical  value   $U_c$ is needed to reach the polarized
ground state. For values of $U<U_c$ the total spin of the ground
state remains  zero. Above the critical value of U the system is
insensitive to the frustrating links and behaves as a normal
bipartite lattice. In the simplest case considered in which we
have a number of vacancies of the same sublattice in the system
and any number of them develop a pentagonal ring, the critical
value obtained to reach the polarized ground state is similar to
the critical U at which the perfect system undergoes phase
transition from the semimetal to an  antiferromagnetic insulator
\cite{ST92,PAB04}.

We are using a single band model for the $\pi$ electrons of graphene
and perform a mean field calculation of  the Hubbard Hamiltonian
$$H=-t\sum_{<ij>,\sigma}c_i^+c_j+U\sum_in_{i\uparrow}n_{i\downarrow},$$
where $<ij>$ stands for nearest neighbors of the honeycomb lattice
and $\sigma$ for the spin polarization. The  tight binding model
for the $\pi$ orbitals is the simplest approach that captures the
electronic structure of graphene \cite{W47} and mean field
calculations of the Hubbard Hamiltonian are often in good
agreement with those obtained by density functional  calculations
in the honeycomb lattice \cite{RMTO02,LMC02}.
\begin{figure}
\begin{center}
\includegraphics[width=7.5cm]{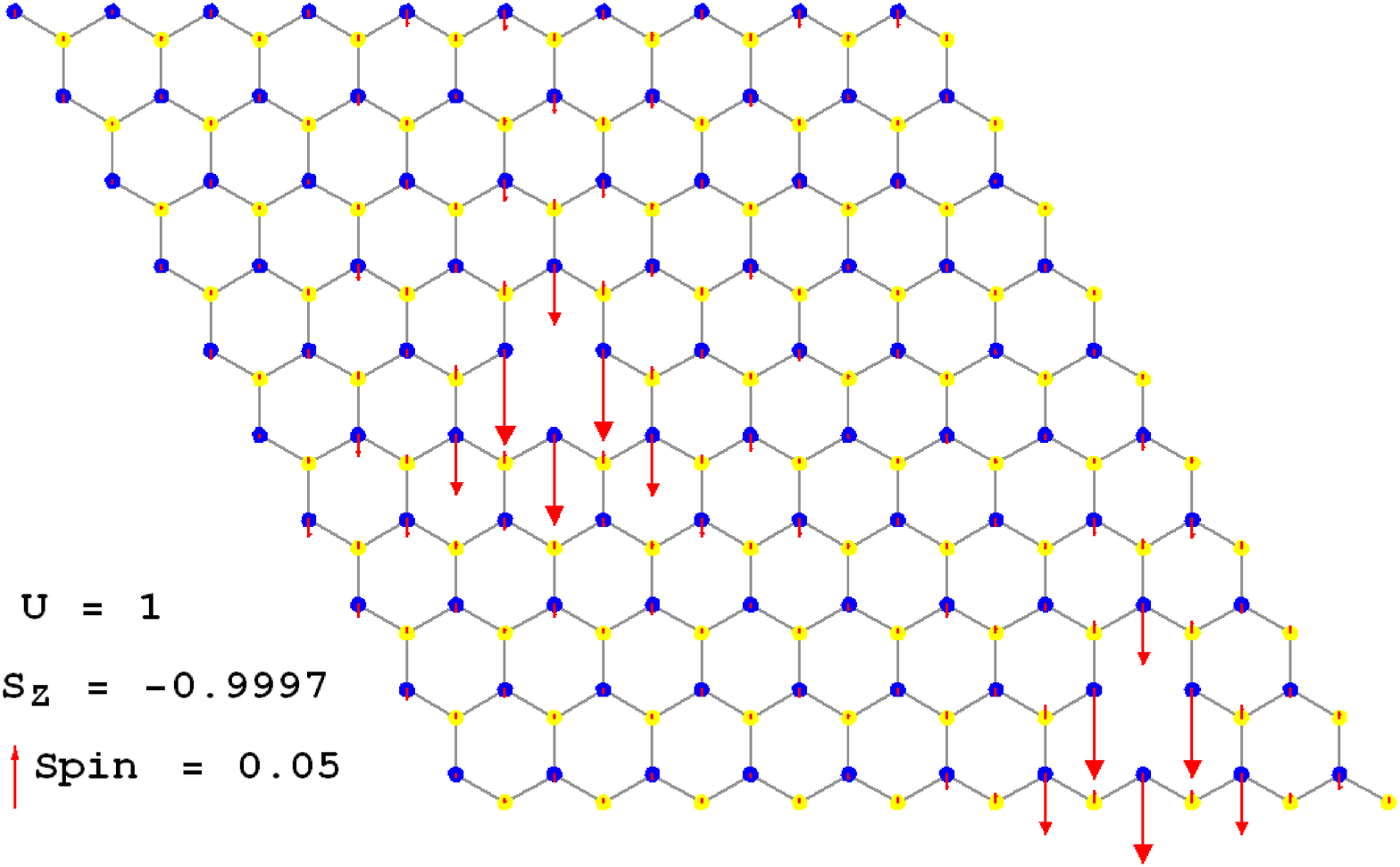}
\includegraphics[width=7.5cm]{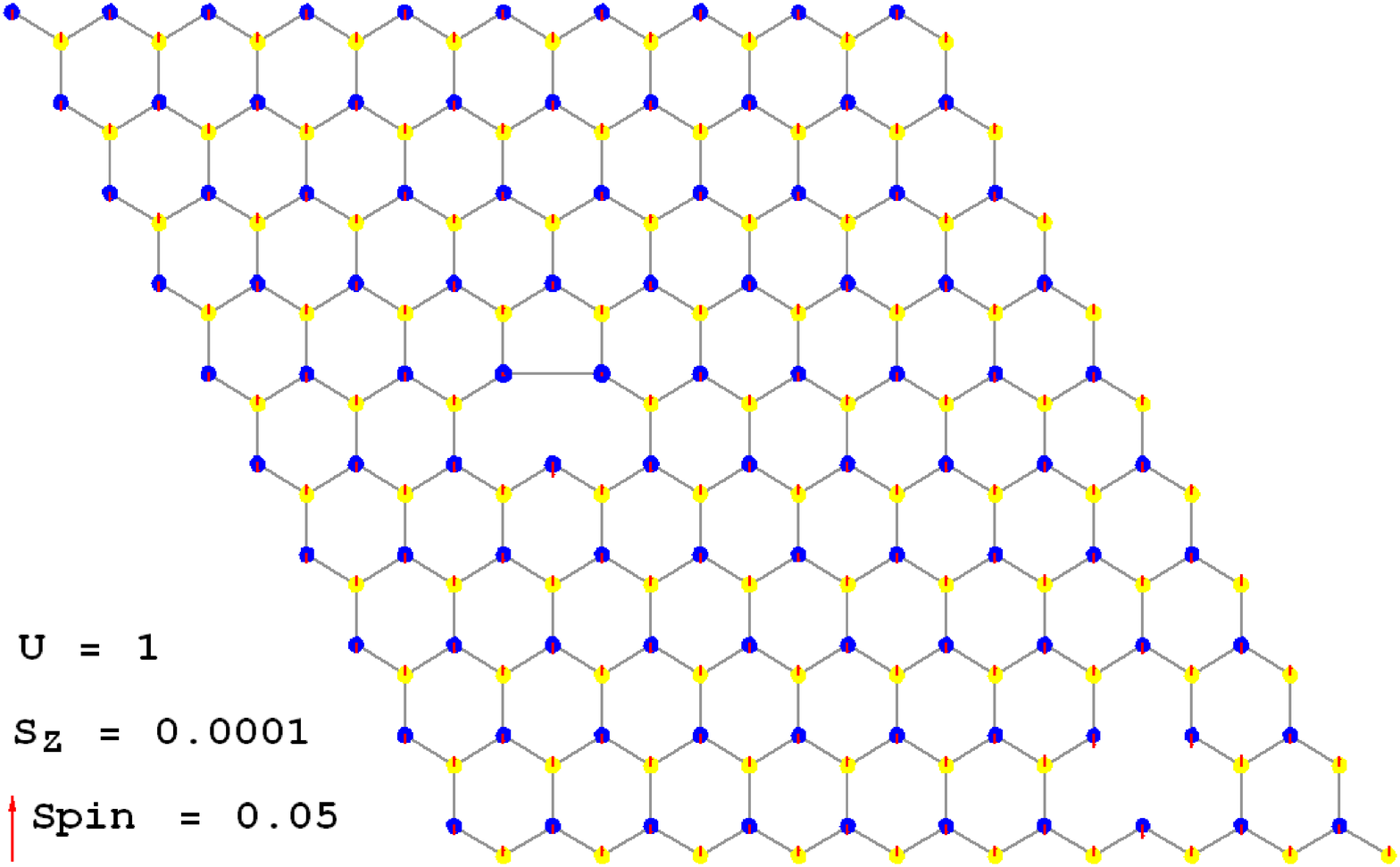}
\caption{Left: Spin distribution in a lattice with two vacancies
of the same sublattice with U=1. Right: Same configuration in the
presence of a pentagon for the same value of U.}
    \label{U1}
\end{center}
\end{figure}
\begin{figure}
\begin{center}
\includegraphics[width=5.5cm]{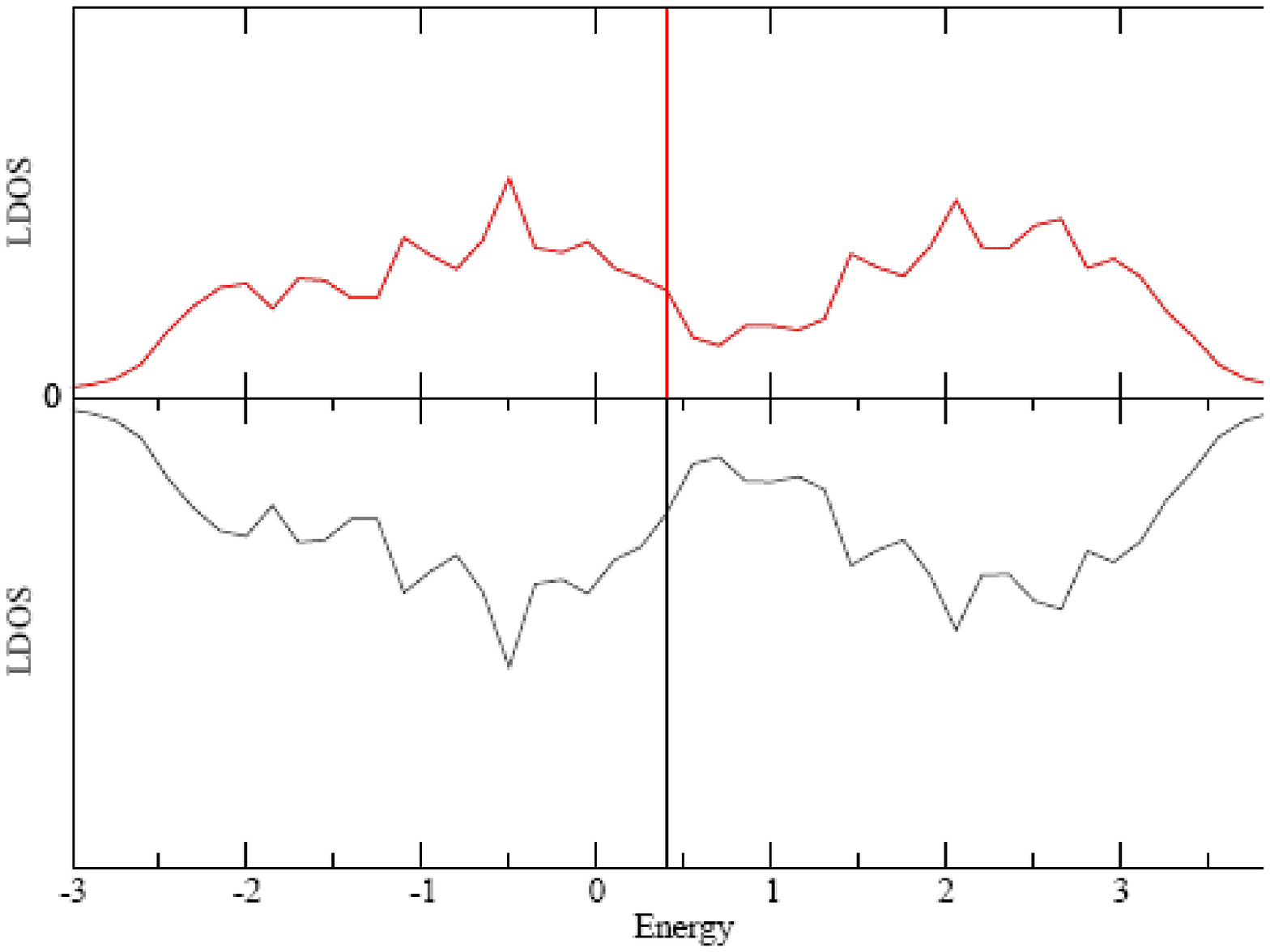}
\includegraphics[width=5.5cm]{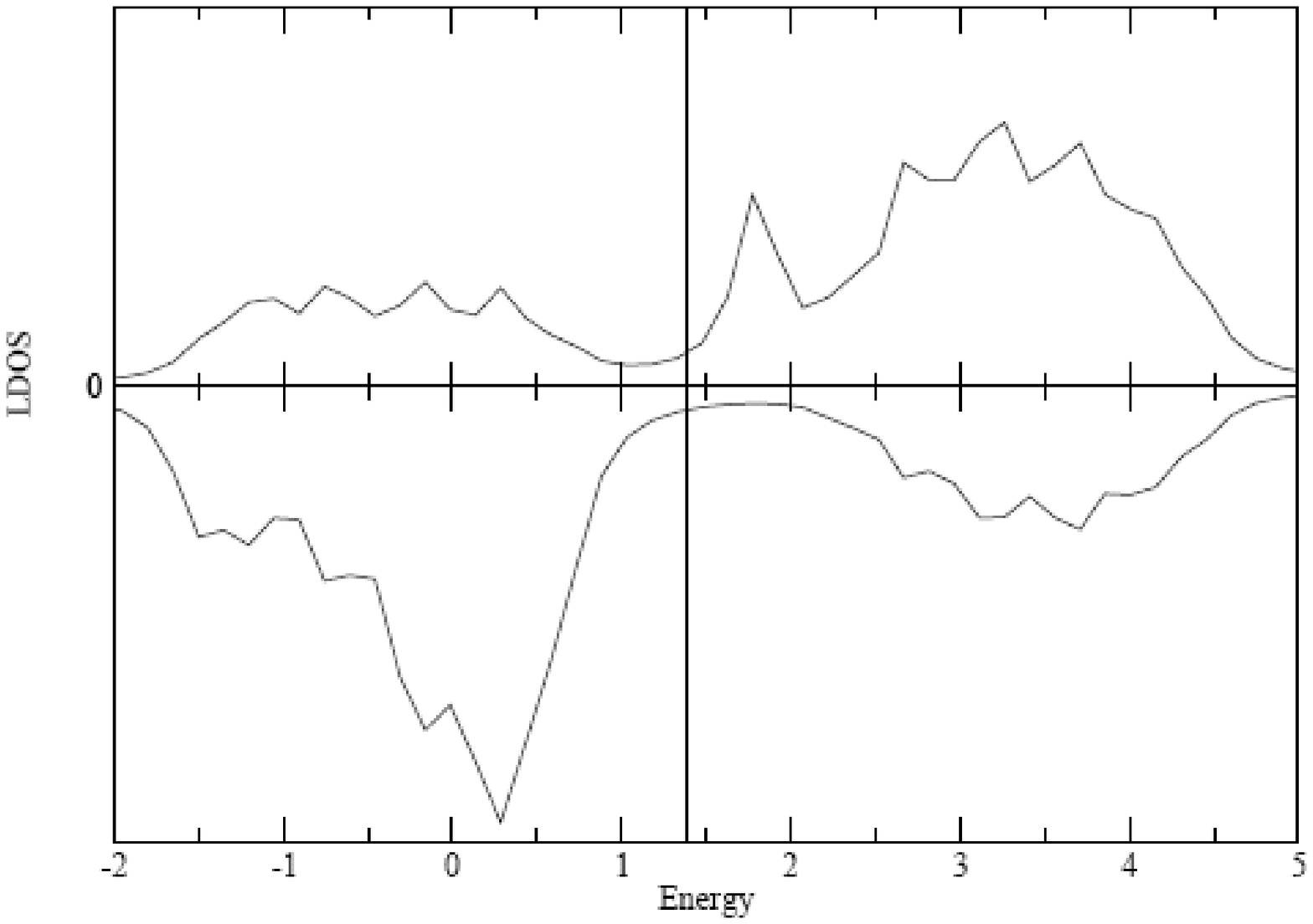}
\caption{Left: Local density of states at the vertex of the
pentagon in the configuration given in Fig. \ref{U1}, right for
the spin up and down electrons with U=1. Right: Same for U=3. The
vertical line shows the position of the Fermi energy.}
   \label{polDOS}
\end{center}
\end{figure}
We begin  studying configurations of two vacancies belonging to
the same sublattice in a graphene sheet where in one of them two
of the unpaired electrons have been joined by a link forming a
pentagon as shown in  Fig. \ref{U1}. This configuration has been
suggested to form naturally as the first step of vacancy
reconstruction \cite{LWetal05,KB07} and has also been shown to
lower the energy in density functional studies of vacancies in
irradiated graphite \cite{LFetal04}. It is the simplest situation
to exemplify the behavior that we want to emphasize.

Fig. \ref{U1} shows the ground state configurations for a value of
the Hubbard repulsion U=1 (throughout the paper U will be measured
in units of the hopping parameter $t$), for both the pentagonal
defect and the vacancy. The total spin of the ground state in the
standard configuration shown in the left side of the figure is
$S_z=1$, what accounts for half the two impaired atoms of the same
sublattice. The polarization for each site of the lattice is
represented by an arrow (its scale in units of $\hbar$ is also
shown adjacent to each figure). We see a relatively strong
polarization localized at the atoms surrounding the vacancy as
expected. In the right hand side of Fig. \ref{U1} one of the
vacancies has relaxed and formed a pentagonal link that we model
with a hopping $t$  of the same value as the rest of the lattice
(this assumption is not important to the results that remain the
same if reasonably different values of the pentagonal $t$ are
assumed). This little frustration of the sublattice order is
enough to destroy the polarization around the two vacancies and
the total spin of the ground state is zero. The structure
presented in the figure corresponds to a density of defects,
vacancies in this case, of one percent which is large. We have
performed the calculation with various defect densities from $0.1$
to $10^{-3}$ and the results remain the same independent not only
of the density of defects but of the relative distances among
them. We have also computed the case in which both vacancies have
a pentagonal link and the results are the same: in the presence of
at least a pentagonal ring there is a critical value of U of
approximately $U\sim 2$ above which the spin of the ground state
recovers the full value $S_z=1$. To better appreciate the effect
of the pentagonal link we note that the critical U to polarize the
ground state for vacancies in the bipartite lattice is zero if the
density of vacancies is not too big. In the non-frustrated case
there is also a transition from an unpolarized semimetal with
magnetic moments strongly localized at the positions of the
uncoordinated atoms surrounding the vacancy to a perfectly ordered
anti-ferromagnetic state with two -frozen- holes and with total
spin determined by the unpaired electrons. The low U configuration
has been described by Lieb in the original paper as an example of
itinerant ferromagnetism, and the high U case as ferrimagnetism,
where there is a perfect antiferromagnetic order in a system with
a non zero total spin. It is quite remarkable that the presence of
a single link frustrating the sublattice symmetry in a cluster or
up to 3200 atoms is enough to rise the critical U to the rather
high value of U=2. As noted before, the critical value found in
this case is similar to the one that sets the
semimetal--AFinsulator transition in the perfect system.

In Fig. \ref{polDOS} we show the local density of states for the
lattice configuration shown at the right hand side of Fig.
\ref{U1}  at a lattice site on the pentagonal defect. The upper
(lower) curve represent the density of electrons with spin up
(down). The left of the figure corresponds to the unpolarized
ground state obtained for a value of U=1 and the right hand side
shows the fully polarized system obtained with U=3. The vertical
line signals the position of the Fermi level which is shifted from
zero by the interaction U. We can see that in the unpolarized
situation the local DOS at the position of the dangling bond is
higher than in the case of the polarized case.

Next we turn to the more interesting case of having dislocations
in the lattice. Recent works on the elasticity in the flat
honeycomb lattice \cite{Cetal08} have demonstrated that two types
of dislocations are stable configurations: shuffle dislocations -
an octagon with a dangling bond- and the more usual glide
dislocations made of a pentagon-heptagon pair. These defects were
described in \cite{EHB02} and experimental observations were
reported in \cite{Hetal04}; dislocations have also been observed
very recently in graphene grown on Ir in \cite{CNetal08}. The
presence of dislocations can affect  the magnetic properties of
the graphene samples in two ways: Shuffle dislocations can
nucleate local magnetic moments similar to the ones induced by
vacancies while the structure of the glide dislocations frustrate
the bipartite nature of the lattice.
\begin{figure}
\begin{center}
\includegraphics[width=7.5cm]{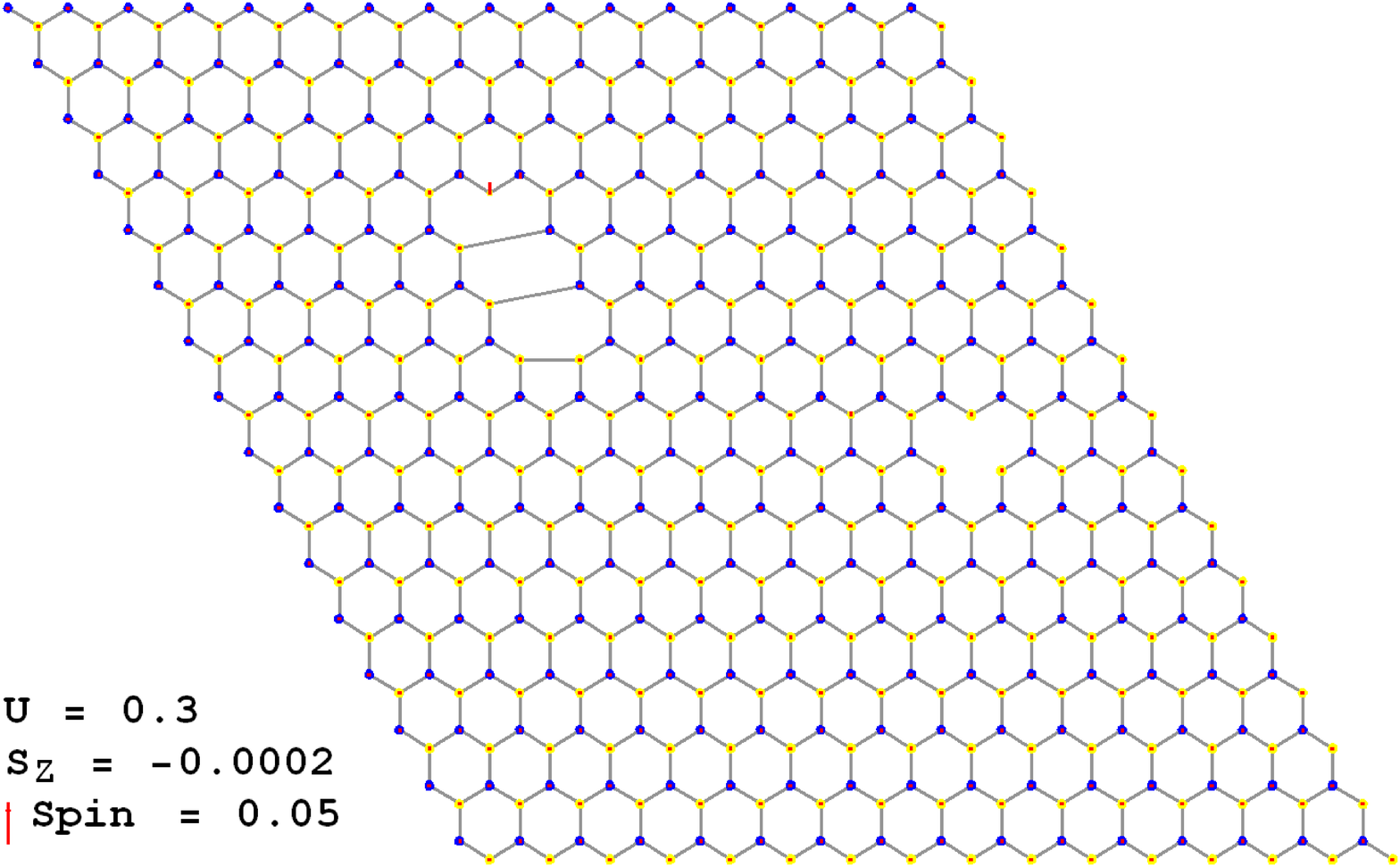}
\includegraphics[width=7.5cm]{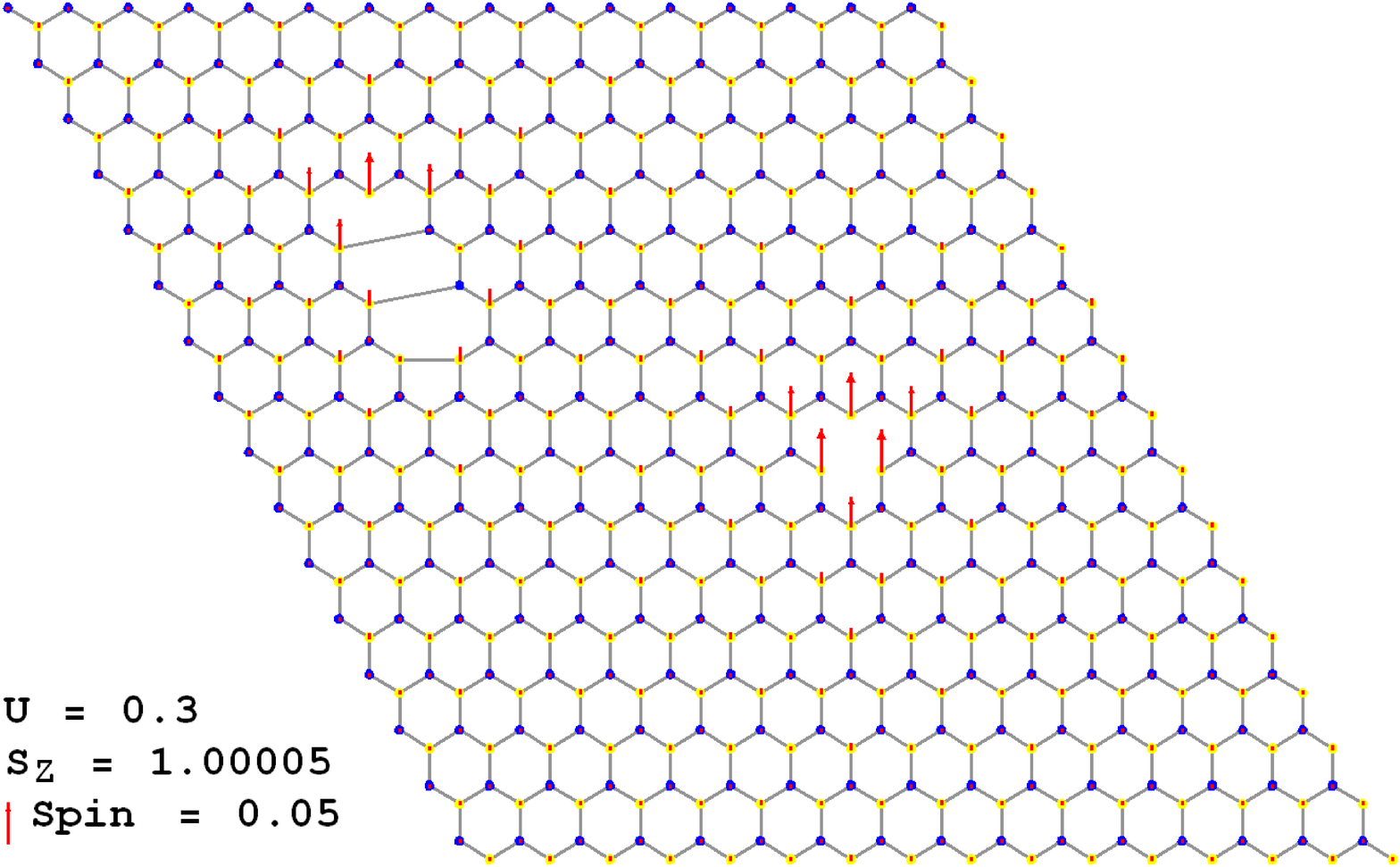}
\caption{Spin structure for two different configurations of
dislocations and a vacancy with U=0.3 with total spin polarization
S=0 (left) and S=1 (right).}
    \label{disloc2}
\end{center}
\end{figure}
Dislocations of either type (glide or shuffle) add -- or
suppress-- a row of atoms to the lattice. In order to eliminate
the influence of the edges and perform the calculation with
periodic boundary conditions we  introduce a pair of dislocations
such that the extra row begins in one and ends in the other one.
Fig. \ref{disloc2} shows the basic structure discussed in this
work. The shuffle dislocation is made of an octagon with an
unpaired atom of a given sublattice. The dislocation line ends in
a glide dislocation made of a pentagon-heptagon pair. This basic
block does not alter the edges of the sample and should behave
like a single vacancy. We have checked that indeed the total spin
of the lattice for this configuration is S=1/2 for a critical
value of $U\sim 0$ showing that the dangling bond of the shuffle
dislocation behaves as a vacancy of the other sublattice, that of
its missing nearest neighbor. If a vacancy of the same sublattice
than the dangling bond atom is added, the total spin of the system
is zero in agreement with Lieb's theorem. When the additional
vacancy belongs to the opposite sublattice, a critical value  of
the interaction $U_c$ is needed to obtain the  total spin S=1. For
$U<U_c$ the total spin is zero. The situation is similar to the
one discussed previously with pentagons but in the case of the
dislocations there is a critical region in the parameter space U
$0.2<U<1$  where the fully polarized and the unpolarized ground
states are almost degenerate in energy  and we find a coexistence
of both cases. In Fig. \ref{disloc2} we show the two spin
configurations obtained at a value of U=0.3 for two defects
located at the same relative distances on the lattice with total
spin $S_z=0$ (left) and $S_z=1$ (right). The critical region
depends on the density and on the relative positions of the
defects and a full phase diagram will be presented elsewhere.

This situation points towards a first order magnetic  transition
in the presence of dislocations but this issue can not be explored
with the techniques of the present work and will be explored in
the future.

We end by a summary of the findings and some remarks. We have
shown that the  nucleation of magnetic moments in the graphene
honeycomb lattice is severely modified by the slightest
frustration of the bipartite character of the lattice. The most
dramatic effect appears when considering standard vacancies of the
same sublattice. It is known that for values of the defect density
not exceeding a certain value of about one percent, the ground
state of the system at half filling has maximal spin given by the
sublattice unbalance for any value of the Hubbard U. This result
has been proven to be quite robust and to apply  for interactions
beyond the Hubbard model. We have shown that the presence of a
single link frustrating the sublattice symmetry in a cluster of up
to 3200 atoms is enough to rise the critical U to a rather high
value of U=2. This critical value is similar to the one that
induces an antiferromagnetic instability in the perfect lattice
estimated to be in mean field of the order of $U\sim 1.8$ although
the value increases when more refined calculations are done up to
$U\sim 4.5$ \cite{ST92}.

As it was explicitly mentioned in the original paper by Lieb
\cite{L89}  the ferromagnetic properties of bipartite lattices
like graphene are determined by the appearance of midgap states
associated to defects and to the electron-electron interactions
within them. The perfect degeneracy of the zero energy states
induced by vacancies or voids belonging to the same sublattice is
broken by the inclusion of a frustrating link and the interplay of
kinetic energy and Coulomb repulsion becomes more subtle. The
importance of the present work relies on the fact that
dislocations and other frustrating defects are very likely to be
present in the real graphene samples and they should be taken into
account.

The findings of this work are somehow negative for the
expectations to get magnetic graphene since the observations of
corrugations in the suspended samples and the high probability of
having dislocations in the mechanically cleaved samples imply the
presence of the topological defects addressed in this work. As a
positive remark we have found that the effect of dislocations
--whose presence we believe to be almost unavoidable in real
samples -- is milder to the magnetism than that of the single
pentagonal rings. The situation is richer and very low critical
values of U are found where the two spin polarizations coexist.
The dependence of the critical U on the density and relative
distances of the defects is being further analyzed and results
will presented elsewhere  but the main aspects discussed in this
paper remain.

We have also found that Stone Wales defects made of two
pentagon-heptagon pairs (two glide disclinations with opposite
Burgers vectors) which are known to play a very important role in
the physics of fullerenes and carbon nanotubes are harmless in the
flat lattice. They were shown to have almost no effect on the
electronic structure \cite{Cetal08} and we have seen that their
presence does not alter the magnetic structure of the unperturbed
lattice. This result is somehow at odds with the effect studied in
ref. \cite{DML04} where Stone Wales defects were assumed to be
responsible for the destruction of the magnetization of graphene
with atomic hydrogen adsorbed but more calculations are needed to
fully explore the issue. Our results can be also of importance in
relation to the recent finding that any edge defect spoils the
metallicity of graphene nanoribbons \cite{EZetal08}.

 \acknowledgments We thank M. J. Calder\'on and A. Ayuela for very useful conversations.
This research was supported by the Spanish MECD grant
FIS2005-05478-C02-01 and by the {\it Ferrocarbon} project from the
European Union under Contract 12881 (NEST).

\bibliography{Lieb}

\end{document}